# Magnetic Bloch States at Integer Flux Quanta Induced by Super-moiré Potential in Graphene Aligned with Twisted Boron Nitride

Yaqi Ma[1,†], Meizhen Huang[1,†,*], Xu Zhang[2,3,†], Weixiong Hu[4], Zishu Zhou[1], Kai Feng[2], Wenhui Li[4,5], Yong Chen[1], Chenxuan Lou[1], Weikang Zhang[1], Haoxi Ji[1], Yibo Wang[1], Zefei Wu[1], Xiaodong Cui[2], Wang Yao[2], Shichao Yan[4,5], Zi Yang Meng[2,*], Ning Wang[1,*]

[1]Department of Physics and Center for Quantum Materials, The Hong Kong University of Science and Technology, Hong Kong, China

[2]Department of Physics and HK Institute of Quantum Science & Technology, The University of Hong Kong, Hong Kong, China

[3]Department of Physics and Astronomy, University of Ghent, 9000 Ghent, Belgium

[4]School of Physical Science and Technology, ShanghaiTech University, Shanghai, China

[5]ShanghaiTech Laboratory for Topological Physics, ShanghaiTech University, Shanghai, China

[†]These authors contributed equally: Yaqi Ma, Meizhen Huang, Xu Zhang

*e-mail: phwang@ust.hk; zymeng@hku.hk; mhuangai@connect.ust.hk

**Two-dimensional electron systems in both magnetic fields and periodic potentials are described by Hofstadter butterfly, a fundamental problem of solid-state physics. While moiré systems provide a powerful method to realize this spectrum, previous experiments, however, have been limited to fractional flux quanta regime due to the difficulty of building ~ 50 nm periodic modulations. Here, we demonstrate a super-moiré strategy to overcome this challenge. By aligning monolayer graphene (G) with 1.0° twisted hexagonal boron nitride (t-hBN), a 63.2 nm bichromatic G/t-hBN super-moiré is constructed, made possible by exploiting the electrostatic nature of t-hBN potential. Under magnetic field $B$, magnetic Bolch states at $\phi/\phi_0 = 1 - 9$ are achieved and observed as integer Brown-Zak oscillations, expanding the flux quanta from factions to integers. Theoretical analysis reproduces these experimental findings. This work opens new avenues to study unexplored Hofstadter butterfly, explore emergent topological order at integer flux quanta and engineer long-wavelength periodic modulations.**

The Hofstadter butterfly is a self-similar recursive energy spectrum to describe the quantum behavior of an electron moving in both a two-dimensional (2D) periodic potential and a magnetic field[1], which is a fundamental problem of solid-state physics. At commensurate fields, electrons recover delocalized wave functions and behave as magnetic Bloch states that propagate as if they are effectively in zero magnetic field. Mathematically, this occurs for all rational values of magnetic flux $\phi/\phi_0 = p/q$, where $\phi = BS$, $S$ is the area of one unit cell, $\phi_0 = h/e$ is the flux quantum, $p$ and $q$ are co-prime integers[2,3]. However, experimental realization of the Hofstadter butterfly has been demanding, because typical atomic lattices process wavelengths in the angstrom scale, requiring unfeasibly large magnetic fields to reach commensurability condition. The rapid development of stacked moiré superlattices[4,5], on the other hand, paves the way to create artificial lattices with periods of tens of nanometers and facilitates the exploration of this effect, in which plenty of significant physics was discovered. In particular, fractional Brown-Zak (BZ) oscillations[6-10], third-generation Dirac points[7], replica quantum Hall ferromagnetism[11] and fractional Chern insulators[12] were observed in the magnetic Bloch states that reside at fractions of $\phi/\phi_0 = 1/q$ and $p/q$ with $p < q$.

However, such fractional sequences of $\phi/\phi_0$ do not constitute the complete Hofstadter butterfly, and the full set of it can appear only when the magnetic Bloch states at integer flux quanta ($\phi/\phi_0 = p$) are reached, which is less explored and has awaited experimental confirmation. The difficulty is constructing much larger moiré wavelengths in the order of 50 nm. In heterostructures, unit cell sizes are limited[4,13], way below the desired length scales. Such length scales are possible in homobilayers at marginal twist, but structural relaxation in reality lead to strong inhomogeneity[14,15]. Until now, the magnetic Bloch states at $\phi/\phi_0 = p$ have only been achieved in limited systems, such as artificially patterned lattices[16,17], twisted bilayer graphene at second magic angle (~ 0.5°)[18], and twisted trilayer graphene[19]. In these cases, problems like charge trap centers and disorders may occur in patterned lattices, while angle inhomogeneity and interlayer coupling induced electronic band changes could appear in twisted graphene systems, all of which could hinder the experimental realization of magnetic Bloch states at $\phi/\phi_0 = p$. Moreover, the achievement of integer flux quanta is the prerequisite to observe some other intriguing states, such as three-dimensional (3D) topological insulator phases emerging exactly at $\phi/\phi_0 = p$ [20] and the reentrant Hofstadter phases[21,22] where the states appear at zero flux could recur at $\phi/\phi_0 = p$, all of which not only contributes to understanding the mechanisms behind the topological phases in the fractal Hofstadter spectrum of moiré bands but also may pave a new route towards practical application prospect, like designing next-generation novel quantum matter and developing fault-tolerant quantum computing.

Here, we construct a super-moiré structure to overcome these difficulties and report the realization of magnetic Bloch states at integer flux quanta. By placing monolayer graphene on a 1.0° t-hBN layers and aligning the graphene with its adjacent hBN layer, two moirés with small unit cells are created, one is 14.4 nm t-hBN, and the other is 13.0 nm G/hBN. The two moirés interfere with each other and form a super-moiré with a large wavelength of 63.2 nm, confirmed by the appearance of satellite peaks near the main Dirac point (MDP). Under magneto-transport, the magnetic Bolch states at $\phi/\phi_0 = 1 - 9$ are achieved, which are observed as integer BZ oscillations with periodicity in $B$, expanding the flux quanta from fractions to integers. In contrast, such BZ oscillations are absent under a single moiré potential with similar wavelength (graphene on 0.23° t-hBN layers), due to the serious moiré potential inhomogeneity. Theoretical analysis quantitatively reproduces the emergence of the super-moiré potentials and the integer BZ oscillations. This work not only opens an avenue to study the Hofstadter system at high flux of $p/q > 1$ in 2D superlattices but also provides a strategy to create high-quality large-period moiré potentials.

## Results
### Construction of super-moiré structure
Commonly used hBN crystals are naturally stacked in an AA' sequence. By stacking two odd-layer AA' hBN films together with a small twist angle $\theta_1$ (corresponding wavelength $\lambda_1$), a parallelly stacked interface can be created. Such stacking leads to a moiré at the interface (t-hBN in Fig. 1a), where the broken inversion and mirror symmetries produce a periodic potential, which is electrostatic and can impact adjacent layers not in direct contact[23-25]. In addition, when a monolayer graphene is further aligned with the bottom layer of t-hBN with an angle $\theta_2$ (related wavelength $\lambda_2$), the lattice mismatch between graphene and hBN gives rise to another moiré (G/hBN in Fig. 1a). The potentials at t-hBN and G/hBN interface process different amplitudes and tunable periods (orange and blue curves as schematically shown in Fig. 1b). In analogous to doubly-aligned hBN/G/hBN superlattices[26-35], by controlling the angle in G/hBN interface to match with the twist angle in t-hBN layers, these two single moirés can interfere with each other, engendering a G/t-hBN super-moiré (middle in Fig. 1a) with a wavelength $\lambda_{sm}$ much larger than that of the individual moirés. Accordingly, electrons in graphene experience a bichromatic super-moiré potential[36] (magenta curve in Fig. 1b), whose period and amplitude are outlined by the black dashed envelope, from the interplay between the G/hBN moiré potential and the electrostatic one at the t-hBN interface. Figure 1c is the counterpart of Fig. 1a in reciprocal space. $\mathbf{G_1}$ and $\mathbf{G_2}$ are the reciprocal lattice vectors of t-hBN and G/hBN single moirés, respectively. The difference between the two vectors gives rise to the reciprocal vector $\mathbf{G}_{sm}$ of the super-moiré, whose Brillouin zone is shown as the small magenta hexagon. Theoretically, super-moiré could form when $\mathbf{G_1}$ and $\mathbf{G_2}$ are commensurate, which happens at specific twist angles (Fig. 1d). (for details, see Supplementary Note 1), while experimentally, super-moiré within a finite sample size could be created approximately when there are specific small deviations from commensurate condition (see discussion in Supplementary Note 2).

We fabricated a device D1 based on the super-moiré structure illustrated in Fig. 1a, where the thickness of hBN flake is about 2 nm (5 layers). Figure 1e shows the schematic diagram of our devices. The t-hBN substrates were assembled by using the "tear and stack" method[5]. Since the electrostatic potential requires a parallelly stacked interface, odd-layer hBN sheets were selected with the second harmonic generation (SHG) technique and then were twisted by a small angle $\theta_1 = 1.0°$ (for even-layer hBN flakes, twist angle is $\theta_1$ plus 60°). The monolayer graphene was then aligned with bottom layer of t-hBN and was picked up. The alignment between graphene and hBN was confirmed by the Raman spectroscopy[37]. The bottom hBN was intentionally misaligned with graphene to avoid additional moiré potentials acting on graphene charge carriers (fabrication details can be found in Methods and Supplementary Note 3). The total charge density $n_{tot}$ induced by electrostatic gating is calculated from $n_{tot} = 1/e[c_{TG}(V_{TG} - V_{TG,0}) + c_{BG}(V_{BG} - V_{BG,0})]$. Here $c_{BG}$ ($c_{TG}$) is the back-gate (top-gate) capacitance, and $V_{BG}$ ($V_{TG}$) is the back-gate (top-gate) voltage. $V_{BG,0}$ and $V_{TG,0}$ are the voltage offsets from back and top gate due to channel impurities, respectively.

Figure 1f shows the measured longitudinal resistance $R_{xx}$ in zero magnetic field as a function of $n_{tot}$ at 1.5 K for device D1. Besides the usual resistance peak from the MDP, two resistance peaks, indicated by the two blue arrows, appear at two sides of the MDP symmetrically, with $n_{tot} = \pm 2.41 \times 10^{12} cm^{-2}$, indicating a moiré potential with wavelength of ~ 13.8 nm. Since the moiré wavelengths generated by $\theta_1 = 1.0°$ t-hBN (14.4 nm) and the aligned G/hBN with $\theta_2$ smaller than 0.5° (12.5 ~ 13.9 nm)[6-8] are very close, the observed two resistance peaks most likely result from overlapped and broadened full fillings of both G/hBN and t-hBN potentials. More importantly, the extra small peaks near the MDP and satellite peak at hole side (indicated by magenta arrows) suggest a potential with a much larger wavelength. The inset is the enlarged plot, in which the magenta curve is the second derivative of $R_{xx}$, $d^2R_{xx}/dn_{tot}^2$, aiming to show the small peaks more clearly. The experimentally measured density of $n_{tot}^0 = 9.07 \times 10^{10} cm^{-2}$ gives rise to a lattice constant of ~ 61.8 nm. From a view of theoretical calculation, such a long wavelength cannot be created by neither G/hBN heterostructure nor 1.0° t-hBN homostructure independently. Instead, it could be a G/t-hBN super-moiré resulting from the interference between the two individual moirés with $\theta_1 = 1.0°$ ($\lambda_1 = 14.4\ nm$), $\theta_2 = 0.4°$ ($\lambda_2 = 13.0\ nm$) and $\lambda_{sm} = 64.6\ nm$ as indicated by the red box in Fig. 1d. As these parameters are consistent with our device fabrication procedure, it is very clear that G/t-hBN super-moiré is constructed in our sample successfully and results in those small satellite peaks in longitudinal resistance. Below we demonstrate that this super-moiré potential can induce magnetic Bloch states at integer flux quanta.

**Magneto-transport under high fields**

The magneto-transport data of device D1 is shown in Fig. 2a, in which the diagram for $R_{xx}$ is plotted as a function of magnetic field $B$ and carrier density $n_{tot}$ at $T$ = 142 K (for low-$T$ data, see Supplementary Fig. 11). At this temperature, Shubnikov-de Haas (SdH) oscillations and the corresponding Landau fans are highly suppressed because of thermal smearing[6-8] and only the $\nu = \pm 2$ Landau fans remain (indicated by the black arrow). Instead, magnetoresistance oscillations that are independent of $n_{tot}$ persist and dominate, i.e., horizontal streaks marked by the white arrows, indicating clear BZ oscillations[2,3,8,9,19]. The inset in Fig. 2a is plotted with the same data but in a narrower $n_{tot}$ range and brighter color filling to highlight the oscillating features. From the positions of such streaks, it is derived that the local maxima in $R_{xx}$ appear periodically in $B$ and locate at a series of value $\phi/\phi_0 = p$ with $p = 1 - 9$. These are integer BZ oscillators, distinct from the Aharonov–Bohm oscillations[38,39] and the fractional BZ oscillations periodic in $1/B$ [6-10]. To better visualize the integer BZ oscillations, we plot the second derivative of $R_{xx}$, $\Delta R_{xx} = \partial^2 R_{xx}/\partial B^2$ as a function of normalized magnetic field $\phi/\phi_0$ and $n_{tot}$ (Fig. 2b). The differentiation procedure effectively removes the smooth background and emphasizes the features by sharpening local maxima ($p$ up to 7 to eliminate the obscure of BZ oscillations caused by $v = \pm 2$ SdH oscillations at high fields). Figure 2c is a plot of $\Delta R_{xx}$ as a function of $\phi/\phi_0$ at fixed $n_{tot} = 4.15 \times 10^{12} \, cm^{-2}$, which is a line cut from Fig. 2b, indicated with the blue arrow. From Fig. 2b and 2c, we see clearly that the positions of $\Delta R_{xx}$ peaks are exactly located at $\phi/\phi_0 = p$, corresponding to magnetic Bloch states at integer flux quanta. From the diagram data, we extracted the oscillation period $\Delta B \sim 1.20 \, T$, which yields the period $\lambda_{sm} \sim 63.2 \, nm$ through $\Delta B S = \phi_0$ and $S = \sqrt{3}/2 \lambda_{sm}^2$, mathematically consistent with the calculated super-moiré wavelength, i.e., 64.6 nm as marked with red box in Fig. 1d. Moreover, the period obtained from the integer BZ oscillations agrees well with the value (61.8 nm) extracted from the positions of small satellite peaks in the longitudinal resistance curve (Fig. 1f). The oscillations measured in different parts of device D1 exhibit similar behaviours (see Supplementary Fig. 12). We note that the existence of two single moirés in one device is not sufficient to generate integer BZ oscillations[40], in contrast, our 63.2 nm G/t-hBN super-moiré resulting from the quantum interference of two parent moirés is sufficient.

To understand how the super-moiré potential produces or facilitates the emergence of magnetic Bloch states at $\phi/\phi_0 = p$, we fabricated another device D2. In device D2, the monolayer graphene was placed on and intentionally misaligned with 0.2° t-hBN layers (thickness about 2 nm, 5 layers, the same in device D1), which is designed to generate a single moiré potential with similar wavelength to that in device D1, as illustrated in Fig. 3a insets. Figure 3a shows the longitudinal resistance $R_{xx}$ measured as a function of $n_{tot}$ at zero magnetic field for device D2 at 1.5 K. Besides the usual resistance peak at MDP, small satellite peaks appear near the MDP, indicated by the orange arrows. The measured carrier density $n_{tot}^0 = 8.88 \times 10^{10} cm^{-2}$ indicates a moiré wavelength of ~ 62.5 nm or a twist angle of ~ 0.23°, consistent with the period extracted from the microscopy images (see Supplementary Fig. 13). The magneto-transport data obtained at 75 K is shown in Fig. 3b (for low-$T$ data, see Supplementary Fig. 11), in which the usual Landau fans characterizing the SdH oscillations can be seen, but no features of BZ oscillation appear, neither with $\phi/\phi_0 = 1/q$ nor $\phi/\phi_0 = p$. We address that the wavelength of potential in device D2 is almost the same as that of the super-moiré device D1, however, the observations in them are different.

**Factors to generate integer BZ oscillations**

Since periodic potential is the prerequisite to generate the Hofstadter butterfly, to confirm the origin of the integer BZ oscillations observed in device D1 and explain its absence in device D2, it is necessary and straightforward to consider the two factors: (i). Potential strength and (ii). Potential periodicity. First, the BZ oscillations could appear only when the periodic modulation is strong enough[41,42]. As illustrated in Fig. 1b, the super-moiré potential could be stronger than the two composition potentials, so it is possible that the t-hBN potential itself cannot produce the BZ oscillations. To demonstrate this, we model the system with a continuous Hamiltonian and compute the conductance with the formula[41,42] (Supplementary Note 1),

$$\sigma_{xx} \propto \int d^2k \sum_n v_n^2 \frac{\partial f^0}{\partial \epsilon_n}$$

Where $f^0$ is the equilibrium Fermi-Dirac distribution function, $\epsilon_n$ is the n-th eigenenergy and $v_n$ is the average of velocity under n-th Hamiltonian eigenstate. The calculation was conducted under various t-hBN potential strength ($V^0_{t-hBN} = 10 - 50\ meV$), and the results with $V^0_{t-hBN} = 20\ meV$ are shown in Fig. 4 while the others can be found in Supplementary Fig. 6. Figure 4a is the calculated conductance $\sigma_{xx}$ plotted with varying $B$ and $n_{tot}$ under G/t-hBN super-moiré potential (device D1). The equally spaced horizontal streaks corresponding to the maxima in $\sigma_{xx}$ can be identified with $\phi/\phi_0 = 1 - 10$, which is in good agreement with the experimental data (Fig. 2) and suggest the observed oscillations originate from the repetitive formation of magnetic Bloch states at $\phi/\phi_0 = p$. Besides, high-order features at other rational $p/q$ values can also be seen but with much weaker amplitudes, which are not observed in device D1 experimentally, this is because their appearance may require stricter conditions, such as ultrahigh doping[9,10]. In addition, the $\sigma_{xx}$ mapping under 0.23° t-hBN potential (device D2) is computed (Fig. 4b and Supplementary Fig. 6). The simulation results with different $V^0_{t-hBN}$ exhibit similar features and reveal three findings. First, the generation of BZ oscillations requires the potential strength to be larger than the energy gap between adjacent Landau levels. Since the gap between Landau levels grows with $B$, for a fixed $V^0_{t-hBN}$, the oscillating features would become dim at high fields (In Fig. 4b, the oscillations at $\phi/\phi_0 = 2, 3$ are visible but faint at higher $B$). Second, with increasing $V^0_{t-hBN}$, the oscillations become more pronounced. Third, as weak as $V^0_{t-hBN} = 10\ meV$, the BZ oscillations start to appear. In reality, the potential strength of t-hBN is measured as at least ~ 50 meV[15,43], which is expected to generate visible BZ oscillations (Supplementary Fig. 6c), even more pronounced than in Fig. 4b. However, no BZ oscillations are observed in our transport data (Fig. 3b). This contradiction suggests that potential strength might not be the dominant factor in our work, and we need to further consider the second factor, potential periodicity.

The potential periodicity directly determines the period of integer BZ oscillations by the relation $\Delta B = \phi_0/S$. To stabilize the BZ oscillations, the potential is desired to process a well-defined periodicity across the device size, otherwise, the oscillating components from each unit cell would have different periods and intercross with each other. In moiré system, the potential periodicity is just the moiré pattern periodicity. Recalling that moiré patterns in marginally twisted bilayers (< 0.5°) are likely to be inhomogeneous and irregular[14,15], the probability is that the potential inhomogeneity leads to the absence of integer BZ oscillations in device D2. To confirm this, we fabricated separate samples (S1, S2 and S3) and conducted the scanning tunneling microscopy (STM) measurements to visualize the moiré patterns directly. In ~ 0° G/hBN, the moiré patterns are homogeneous and almost the same with period of 13.4 nm at different positions of the sample (Supplementary Fig. 14a, S1), consistent with the fact that the moiré pattern in heterostructures is dominated by the lattice mismatch and does not change much with twist angle. Similarly, the moiré patterns of ~ 1° t-hBN are uniform and regular (Supplementary Fig. 14c, S2), despite the slight differences in wavelength of 10.2 – 11.5 nm. In contrast, in marginally t-hBN (Supplementary Fig. 14e, S3), the moiré patterns are constituted by triangular domains with large size differences of ~ 98.2 – 167.3 nm, exhibiting explicit inhomogeneity and irregularity, similar results were reported in previous studies[14,15].

This difference can be understood by noticing structural relaxation at small twist angles, which can increase the area of certain energetically favorable stacking while shrink the unfavorable ones, at the cost of inhomogeneous strain[14,44], and for marginal twist (< 0.5°), even minor strain variations lead to large changes in the moiré wavelength ($\lambda \propto 1/\theta$). In general, as the size of domains increases, the moiré pattern is more likely to become irregular. To be more specific, the local twist angle variation in magic angle (1.1°) twisted bilayer graphene was investigated up to 0.1°[45], if we assume this variation is the same in ~ 1° and ~ 0.2° t-hBN, then the consequent moiré periods are in the range of 13.0 – 15.9 nm and 48 – 143 nm, respectively. Thus, it is reasonable that the G/t-hBN super-moiré potential constructed with G/hBN and ~ 1° t-hBN tends to be much more homogeneous and regular, compared to ~ 0.2° t-hBN, although the potential inhomogeneity occurs in both systems. Further, the poor potential periodicity could be the primary reason for the absence of BZ oscillations in device D2, instead of the potential strength. Overall, our theoretical calculation and STM results explain the transport observations well. The integer BZ oscillations appear when the G/t-hBN super-moiré potential is uniform and strong (Fig. 2 and Fig. 4a) but disappear when the 0.23° t-hBN moiré potential is inhomogeneous (Fig. 3b and Supplementary Fig. 14).

## Discussion

Traditional super-moiré structures, such as hBN/G/hBN[26-35] and twisted trilayer graphene[19,46-50], are composed of the same type of parent moirés. The super-moiré is formed through the proximity coupling[51], which dominates the modulation strength, so the potential strength is less tunable. Furthermore, due to the 2D nature of this structure, it is almost impossible to incorporate more a third single moiré to form a high order super-moiré. In contrast, in G/t-hBN super-moiré, which consists of different types of individual moirés, t-hBN potential originates from the electrical polarization due to broken inversion and mirror symmetries[23], so the electrostatic nature of it allows for the spatial separation of the two single moiré interfaces, facilitating to tune the potential strength by varying the thickness of hBN. Moreover, G/t-hBN structure overcomes the limitations imposed by the 2D structure, which enables the inclusion of more single moirés to generate high order super-moirés. In addition, the interlayer couplings in twisted trilayer graphene could highly change the electronic bands, which is likely to compete with and overwhelm the features of interest. Differently, t-hBN acts as a substrate and is not involved in electron transport directly, helping G/t-hBN super-moiré to be a relatively clean system. Overall, emerging as a new type of super-moiré, G/t-hBN structure opens new opportunities for designing long-wavelength periodic modulations with electrostatic potential form t-hBN substrate, which holds promising application prospect, like confining excitons for high-quality single-photon emission arrays[52], a long-standing goal in the field of solid-state quantum information technology[53].

In conclusion, to reach the integer flux quanta in Hofstadter system, we established a strategy of moiré interference to engineer the ~ 50 nm periodic potentials. By aligning monolayer graphene with 1.0° t-hBN substrate, a 63.2 nm G/t-hBN super-moiré is constructed, made possible by exploiting the electrostatic nature of the potentials at t-hBN interfaces and overcoming the inhomogeneity difficulty in marginally twisted single moirés due to the structural relaxation. The fascinating consequence is the achievement of magnetic Bloch states at $\phi/\phi_0 = 1 - 9$, which are observed as integer BZ oscillations with periodicity in $B$, successfully expanding the flux quanta from factions to integers. Further, with the Hofstadter spectrum being a unique platform, the effect of topology[20,54] and nonzero Chern numbers associated with magnetic minibands[55] on transport properties could be explored in the future.

## Methods

**Transport device fabrication.** The devices were assembled with the standard dry-transfer method[56] and tear and stack technique[5]. First, the thin hBN crystals and graphite were mechanically exfoliated onto a Si wafer with 285 nm $SiO_2$ layer. Thin hBN (~ 2 nm) flakes and monolayer graphene were then identified by optical microscopy. To fabricate the aligned devices, the sheets (both hBN and graphene) with long straight edges were selected, which indicate their crystallographic axes. The transfer stamp was made from polydimethylsiloxane (PDMS) film covered with a thin layer of polypropylene carbonate (PC). With a precise micromanipulator the PC film was placed on the top of thin hBN, which is on the $SiO_2$ substrate and whose straight edge was marked. The PC film expanded with heating the transfer stage, when it covered approximately half of the hBN flake, the heating was stopped, and the stage was maintained at high temperature (about 80 °C) for several minutes. Then, the hBN crystal was picked up slowly from the substrate by cooling the stage, leading to the hBN layer being divided into two parts (Supplementary Fig. 8a). Later, the remaining hBN part was rotated by a small angle and subsequently was picked up by the former hBN layer, from which the t-hBN substrate was fabricated. Next, the monolayer graphene on $SiO_2$ substrate was placed on the stage, whose straight edge was aligned (misaligned) with that of bottom layer of t-hBN substrate, to fabricate aligned (misaligned) G/t-hBN samples, and then was picked up by the t-hBN substrate in the same process. Finally, the three flakes on PC film were released on a bottom hBN sheet prepared previously, which was intentionally misaligned (Supplementary Fig. 8b). Standard electron-beam lithography processes and reactive ion etching were followed to expose the edge contact regime, where metal deposition was employed subsequently to construct electrical leads consisting of Cr/Au: 5 nm/70 nm. Further electron-beam lithography and metal deposition were used to define the top gate electrode and to shape the stack into Hall bar devices (Supplementary Fig. 8c).

**STM sample fabrication.** The STM samples were assembled with similar processes of fabricating transport devices but had different stack orders. During the transfer process, graphene, hBN and the remaining half of hBN were picked up by PC successively, and then the three-layer stack was placed on the highly ordered pyrolytic graphite (HOPG) substrate.

**SHG measurement.** The second harmonic generation (SHG) process is employed as a sensitive probe to determine the symmetry of layered crystals, which is described by a third order tensor with electric dipole approximation[57]. For thin hBN flakes, the broken inversion symmetry in odd-layer stack leads to a huge enhancement of the SHG signal, while in even-layer configuration, the hBN crystal is centrosymmetric and the SHG response reduces to nearly zero[58]. Thus, the SHG response can be used to determine the odd/even layers of thin hBN crystals. In the SHG measurement, the 800 nm pulse laser from a Mira900 mode-locked Ti: sapphire oscillator, with 120 fs pulse duration and 76 MHz repetition rate, was focused on the sample (thin hBN) through a 50× objective (spot size 3 μm). The SHG signal was then collected by a Charge-coupled detector (CCD) mounted on a spectrometer. Before the reflected signal entered spectrometer, it first went through a 600 short-pass filter to eliminate 800 nm laser, leaving only 400 nm SHG signal to be collected. Supplementary Fig. 9 shows the typical SHG signals of few-layer hBN flakes used in our devices. Supplementary Fig. 9a and 9b are the optical images of hBN layers probed, with thickness of about 2 nm (5 layers) and about 3 nm (10 layers), respectively. It is clear that a strong SHG signal occurs for odd numbers of layers (red curve in Supplementary Fig. 9c), while even numbers of layers do not show a measurable SHG signal (blue curve).

**Raman spectrum.** The 2D peak in Raman spectrum is sensitive to the misalignment between graphene and hBN, which is broadened when the misalignment angle is smaller than 2°[37]. Thus, the broadness of 2D peak can be used as a probe to verify the alignment between graphene and hBN. To fabricate the G/t-hBN super-moiré devices, where monolayer graphene is designed to align with the bottom layer of t-hBN substrate, when the transfer process was completed the Raman spectroscopy was followed to examine the alignment. The Raman spectra were acquired using an inVia Qontor Raman spectrometer with excitation laser wavelength of 532 nm, power of 1 mW, beam spot size of 1 μm and resolution of ~ 1.5 $cm^{-1}$. Supplementary Fig. 10 shows the Raman spectrum data taken from the super-moiré device D1.

**STM measurement**. Although the atomic force microscope (AFM) is often utilized to image the moiré patterns[59,60], here we use the STM to conduct the visualization, due to its much higher spatial resolution. The STM experiments were conducted with a Unisoku ultrahigh-vacuum STM at 4.8 K. The tungsten tips were flashed by electron-beam bombardment for several minutes before use. Typical imaging parameters were sample voltages between 100 mV and 500 mV and tunneling currents between 10 pA and 500 pA.

**Electrical characterization**. Transport measurements were performed in a cryogenic system, which provides stable temperatures ranging from 1.4 to ∼ 300 K and fields up to 14 T. AC bias voltage was applied to the source probe through Stanford Research Systems DS360. The current and voltage were measured with low-frequency lock-in technique (SR 830 with SR550 as the preamplifier).

**Theoretical calculation.** The band structures of graphene under G/t-hBN super-moiré potential (device D1) and t-hBN single moiré potential (device D2) were computed based on a continuum model of Dirac fermion. Under strong magnetic fields, the fractal energy spectrums were solved by exactly diagonalizing the Hamiltonian after the moiré potentials were transformed into Dirac fermion Landau level basis, from which the longitudinal conductance $\sigma_{xx}$ was obtained as a function of normalized magnetic field and filling by using the Diophantine equation and the Drude formula. All the theoretical plots are shown in Fig. 1c, 1d and Fig. 4 in the main text, as well as in Supplementary Note 1.

## Data availability

Relevant data supporting the key findings of this study are available within the article and the Supplementary Information file. All raw data generated during the current study are available from the corresponding authors upon request.

**Acknowledgements**

Grant support from the National Key R&D Program of China (2020YFA0309600) and the Research Grants Council

(RGC) of Hong Kong (Project Nos. AoE/P701/20, 16303720, and C7037-22GF) are acknowledged. X. Z. and Z. Y. M acknowledge the support from the Research Grants Council (RGC) of Hong Kong Special Administrative Region of China (Project Nos. 17301721, AoE/P-701/20, 17309822, HKU C7037-22GF, 17302223), the ANR/RGC Joint Research Scheme sponsored by RGC of Hong Kong and French National Research Agency (Project No. A HKU703/22) and the HKU Seed Funding for Strategic Interdisciplinary Research "Many-body paradigm in quantum moiré material research". X. Z. and Z. Y. M also thank HPC2021 system under the Information Technology Services and the Blackbody HPC system at the Department of Physics, University of Hong Kong, as well as the Beijing PARATERA Tech CO., Ltd. (URL: https://cloud.paratera.com) for providing HPC resources that have contributed to the research results reported within this paper. Device fabrication was conducted at the MCPF and WMINST of HKUST with great technical support from Mr. Chun Kit Lai, Mr. Gordon C T Suen and Dr. Yuan Cai.

## Author Contributions

M. H. and Y. M. conceived and designed the experiments. Y. M. fabricated the devices and performed the transport measurements under the instruction of M. H. and N. W.. X. Z. performed theoretical computations under the supervision of Z. Y. M.. Y. M and Z. Z fabricated the STM samples. W. H and W. L conducted the STM measurements under the supervision of S. Y.. K. F. conducted the SHG measurements under the supervision of X. D. C.. Y. M. and M. H. analyzed the data and wrote the manuscript. N. W., Z. Y. M, W. Y. and X. Z. polished the manuscript. Y. C., C. L. W. Z., H. J., Y. W. and Z. W. provided technical support in the device fabrication process. N. W. finalized the manuscript with contributions from all authors.

## Competing interests

The authors declare no competing interests.

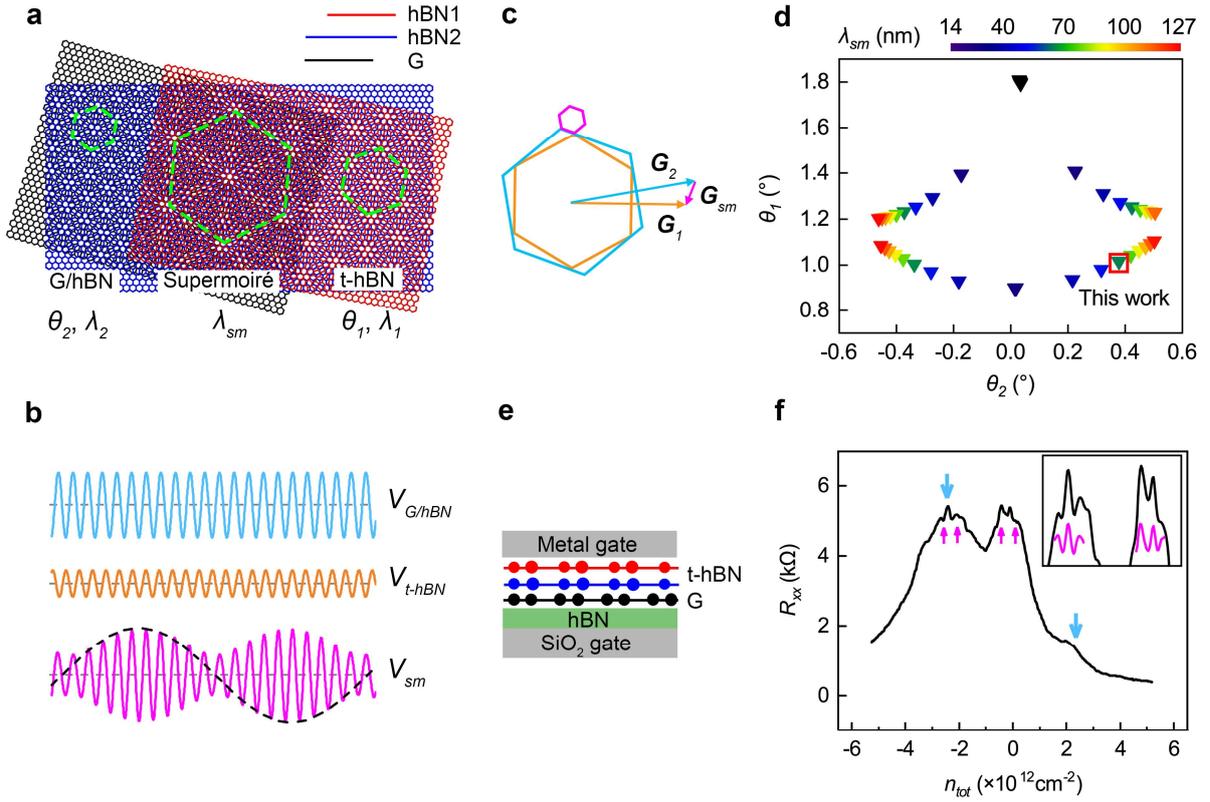

**Fig. 1. Super-moiré potential in device D1**. **a**. Illustration of the two single moirés and super-moiré. The red (blue) indicates the top (bottom) layer of twisted hBN (t-hBN) and black is graphene. The wavelengths of t-hBN, G/hBN and G/t-hBN super-moiré are $\lambda_1 \sim 14.4\ nm$, $\lambda_2 \sim 13.0\ nm$ and $\lambda_{sm} \sim 63.2\ nm$ in device D1, respectively, outlined by three green dashed hexagons. **b**. Schematic illustration of the superimposition of potentials from t-hBN ($V_{t-hBN}$) and G/hBN ($V_{G/hBN}$) layers, which generates a super-moiré potential ($V_{sm}$) with a much larger wavelength. **c**. Schematics for the formation of super-moiré in reciprocal space, which is the counterpart of **a**. $\mathbf{G}_1$, $\mathbf{G}_2$ and $\mathbf{G}_{sm}$ are the reciprocal lattice vectors for t-hBN, G/hBN and G/t-hBN super-moiré, respectively, and the three hexagons are corresponding Brillouin zones. **d**. Calculated $\lambda_{sm}$ plotted as a function of $\theta_1$ and $\theta_2$. Only commensurate super-moiré configurations with wavelengths larger than 14 nm are considered and shown. The super-moiré wavelength realized in this experiment is denoted by the red box, with $\theta_1 = 1.0°$, $\theta_2 = 0.4°$ and $\lambda_{sm} = 64.6\ nm$. **e**. Schematic side view of device D1, where monolayer graphene is aligned with t-hBN substrate. **f**. Longitudinal resistance $R_{xx}$ as a function of carrier density $n_{tot}$ at 1.5 K. Two satellite resistance peaks, indicated by blue arrows, are related to the full fillings of the t-hBN and G/hBN moiré potentials. The small satellite peaks, marked by the magenta arrows, are related to the super-moiré potential. The inset is an enlarged plot, where the magenta curve is the second derivative of $R_{xx}$, $\mathrm{d}^2 R_{xx}/\mathrm{d}n_{tot}^2$.

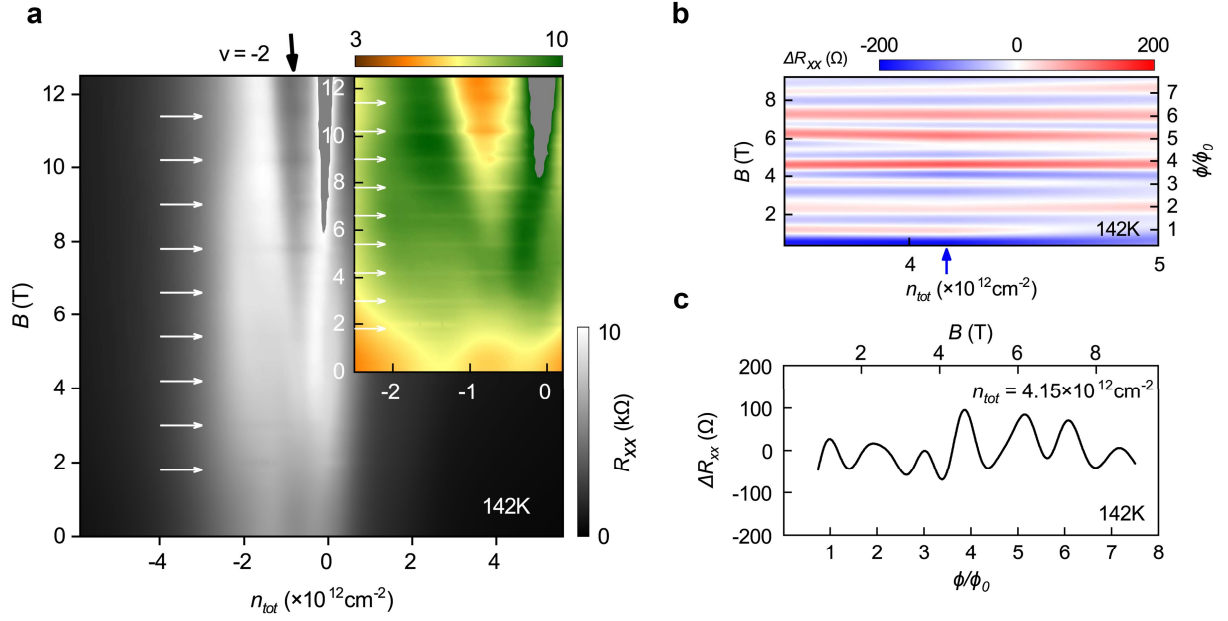

**Fig. 2. Magnetic Bloch states at integer flux quanta induced by super-moiré potential in device D1. a.** $R_{xx}$ as a function of $B$ and $n_{tot}$ at 142 K. The fractal features appear at $B = p(\phi_0/S)$ with $p = 1 - 9$, $S$ the area of unit cell, $\phi_0$ the flux quantum, suggesting the integer BZ oscillations periodic in $B$. Inset is the same data but plotted in a narrower $n_{tot}$ range and brighter color filling. **b.** Second derivative $\Delta R_{xx} = \partial^2 R_{xx}/\partial B^2$ of the same data in **a**, plotted as a function of normalized magnetic field $\phi/\phi_0$ and $n_{tot}$, which removes the smooth background and highlights the oscillating features. **c.** $\Delta R_{xx}$ as function of $\phi/\phi_0$ at fixed $n_{tot} = 4.15 \times 10^{12} cm^{-2}$, which is a line cut from **b** marked by the blue arrow. The integer BZ oscillations periodic in $B$ are clearly seen and the $B$ spacing of $\Delta B = 1.20\ T$ implies a super-moiré wavelength of ~ 63.2 nm.

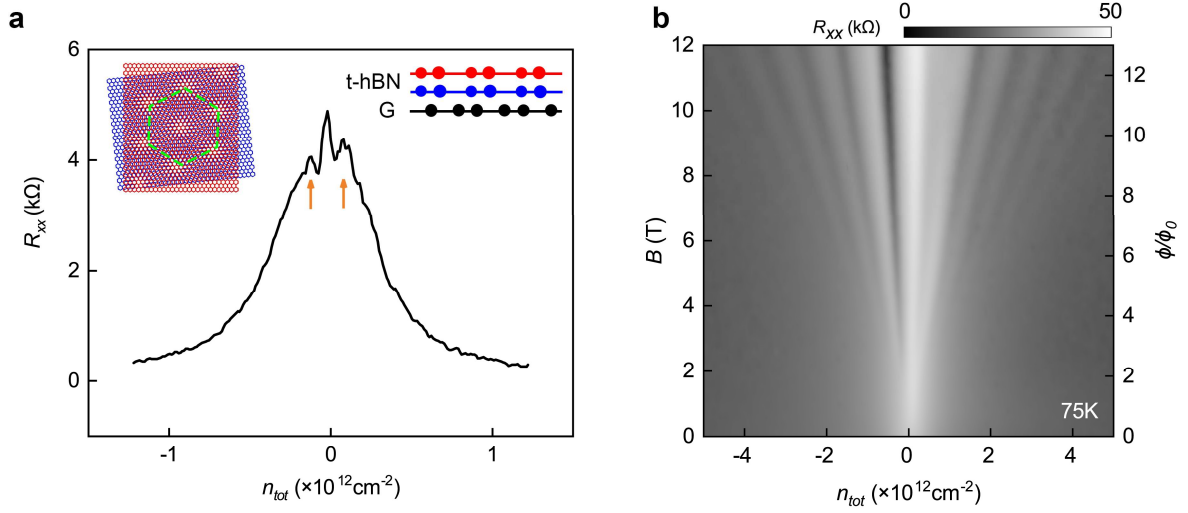

**Fig. 3. Absence of fractal oscillations in 0.23° t-hBN device D2**. **a**. $R_{xx}$ measured versus $n_{tot}$ under zero magnetic field at 1.5 K. The satellite peaks around the main Dirac point (MDP), marked by the orange arrows, are related to the superlattice Dirac cones. From their carrier densities the size of t-hBN unit cell can be calculated as $S = 4/n_{tot}$, giving rise to a wavelength of ~ 62.5 nm. Left inset is the moiré pattern at the t-hBN interface, outlined by the green dashed hexagon. The red (blue) indicates the top (bottom) layer of t-hBN. Right inset is the side view of the stack configuration in device D2, where graphene is misaligned with the t-hBN layers purposely. **b**. Landau fan diagram of device D2 measured at 75 K. Besides the typical Shubnikov-de Haas (SdH) oscillations for monolayer graphene, no oscillation is observed.

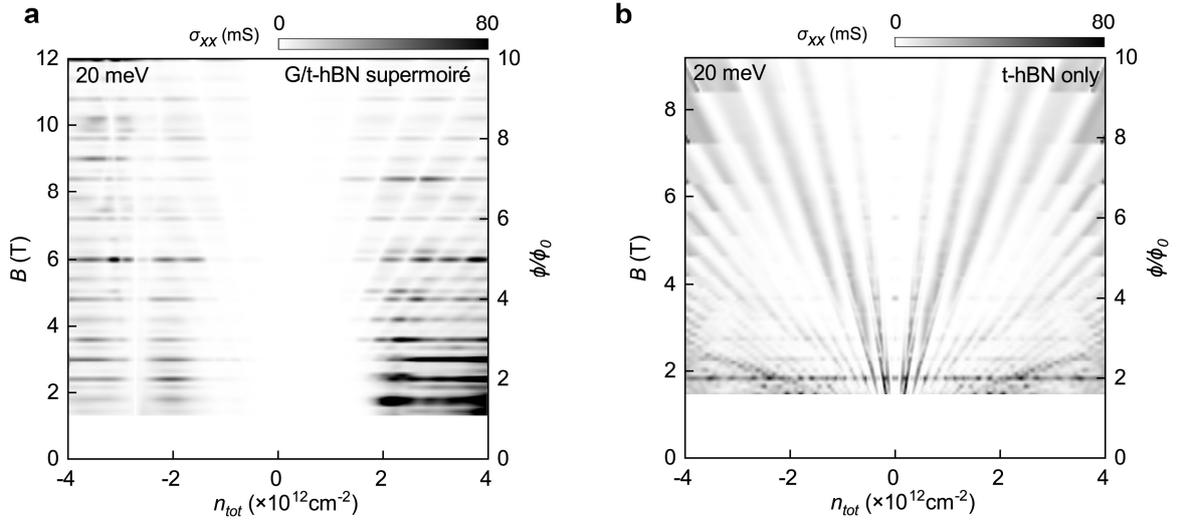

**Fig. 4. Magneto-transport from theoretical calculation**. **a**. Calculated conductance $\sigma_{xx}$ plotted with varying $\phi/\phi_0$ and $n_{tot}$ under super-moiré potential (device D1). The strongest fractal features appear at $\phi/\phi_0 = p, (p = 1, 2, 3, ...)$. The magnetic Bloch states with other rational $p/q$ values ($p$ and $q$ co-prime integers) can also be seen but with much weaker contrast, such as half integers. **b**. $\sigma_{xx}$ mapping computed under 0.23° t-hBN potential (device D2) with $V^0_{t-hBN} = 20\ meV$, the same in **a**. The oscillating features can be seen at low fields but become faint with increasing $B$. In **a** and **b**, the Landau levels are present as the background. Note that the blank at low $\phi/\phi_0$ is due to the limited numbers of Landau levels involved in calculation and hence contain no meaningful data. The details of the calculation can be seen in Supplementary Note 1.